\documentclass[aps,letterpaper,twocolumn,preprintnumbers,floatfix,superscriptaddress]{revtex4}

\usepackage{graphicx}
\usepackage{dcolumn}
\usepackage{bm}
\usepackage{epsfig}
\usepackage{gensymb}
\usepackage{mathrsfs}  
\usepackage{amsmath}
\usepackage{amssymb}
\usepackage{graphicx,bm}
\usepackage{slashed}
\usepackage[makeroom]{cancel}
 \usepackage[titletoc]{appendix}

\def\fun#1#2{\lower3.6pt\vbox{\baselineskip0pt\lineskip.9pt
  \ialign{$\mathsurround=0pt#1\hfil##\hfil$\crcr#2\crcr\sim\crcr}}}

\def\lsim{\mathrel{\rlap{\raise 2.5pt \hbox{$<$}}\lower 2.5pt\hbox{$\sim$}}}
\def\gsim{\mathrel{\rlap{\raise 2.5pt \hbox{$>$}}\lower 2.5pt\hbox{$\sim$}}}

\input epsf

\newcommand{\be}{\begin{equation}}
\newcommand{\ee}{\end{equation}}
\newcommand{\bea}{\begin{eqnarray}}
\newcommand{\eea}{\end{eqnarray}}

\usepackage{color}

\newcommand{\comment}[1]{}

\begin{document}

\title{The Maximally Symmetric Composite Higgs}

\author{Csaba Cs\'aki}
\affiliation{ Department of Physics, LEPP, Cornell University, Ithaca, NY 14853, USA}
\author{Teng Ma}
\affiliation{ 
 CAS Key Laboratory of Theoretical Physics, Institute of Theoretical Physics,
Chinese Academy of Sciences, Beijing 100190, China.}
\author{Jing Shu}
\affiliation{ 
 CAS Key Laboratory of Theoretical Physics, Institute of Theoretical Physics,
Chinese Academy of Sciences, Beijing 100190, China.}
\affiliation{ CAS Center for Excellence in Particle Physics, Beijing 100049, China}
\begin{abstract}

We present a novel class of calculable four dimensional composite pseudo-Goldstone boson Higgs models based on symmetric $G/H$ coset spaces  which contain a Higgs-parity operator $V$ as well as a linear representation $\Sigma'$ for the Goldstone bosons. For such cosets the low-energy  effective Lagrangian for the  Standard Model fields can have an enhanced global symmetry  which we call the maximal symmetry. We show that  such a  maximally symmetric case leads to a finite and fully calculable Higgs potential, which also minimizes the tuning by eliminating double tuning and reducing the Higgs mass.  We present a detailed analysis of the Maximally Symmetric $SO(5)/SO(4)$ model, and comment on its observational consequences. 

\end{abstract}


\maketitle

\section{Introduction}

Electroweak symmetry breaking (EWSB) is a key ingredient of the standard model (SM) responsible for all elementary particle masses. While the discovery of the Higgs 
boson~\cite{Aad:2012tfa,Chatrchyan:2012xdj} was a major milestone towards understanding the mechanism of EWSB, several important issues remain unexplained. In analogy with superconductivity the Higgs potential is assumed to be of the simplest Landau-Ginzburg type~\cite{Ginzburg:1950sr}. In the condensed matter systems we understand that the potential describes the condensation of emergent collective modes, however in particle physics we don't even know if the Higgs is elementary or composite, and what the true Higgs potential is. 
We would also like to understand  whether the Higgs potential is stable under large quantum corrections in the ultraviolet (UV) and whether a small or large fine tuning is needed to maintain the hierarchy. 

The idea that the Higgs is a pseudo-Nambu-Goldstone boson (pNGB)~\cite{Kaplan:1983fs,Georgi:1984af,Dugan:1984hq} of spontaneously broken approximate global symmetry of some strong dynamics gives intriguing answers to the above mysteries. In this scenario, the Higgs could be a bound state of some strongly coupled constituents, while the entire Higgs potential is radiatively generated via loops from the top and gauge sectors, which will trigger vacuum misalignment and EWSB. Due to its pNGB nature, the Higgs mass remains naturally light. The cutoff scale is reduced to the confinement scale $\Lambda \sim 4 \pi f$ (where $f$ is the scale of global symmetry breaking). The sensitivity of the Higgs potential to this confinement scale can be further reduced by different mechanisms~\cite{ArkaniHamed:2001ca,LH,SILH,Bellazzini:2014yua}.  However the parameters of the existing models have to be tuned to achieve a realistic Higgs potential and particle spectrum. The origin of this tuning is to ensure that the EWSB VEV $v$ is small compared to the global symmetry beaking scale $f$, $v/f\ll 1$ to evade electroweak precision~\cite{Peskin:1990zt} and direct detection bounds for the top partners~\cite{Aad:2015kqa,Khachatryan:2015oba}.  

In this paper, we propose a novel type of composite Higgs model that can address the above issues and require only the minimal structure of the general low-energy  Goldstone boson (GB) Lagrangian.  We consider models where $G/H$ is a symmetric space, implying the existence of a Higgs parity operator $V$ as well as a linear representation $\Sigma' = U^2 V$ for the Goldstone bosons. The original symmetry $G$ can be used to easily find the general form of the 
low-energy effective Lagrangian in terms of the GB's using $\Sigma' $ and the SM fields which are assumed to be embedded in (spurionic) full representations of $G$.  In addition, the left- and right-handed components of the SM fermions have an enlarged $G_L\times G_R$ chiral symmetry (in the absence of the GB matrix $\Sigma'$).  Our main observation is that the presence of the SM fermion mass terms (originating from $\Sigma'$) will not completely break the 
$G_L\times G_R$, but rather leave a subgroup $G_{V'}$ (which keeps the pNGB field invariant $g_L \Sigma' g_R^\dag = \Sigma'$) unbroken. The appearance of this ``maximal symmetry" has wide-ranging consequences: the contribution of the top sector to the Coleman-Weinberg (CW) Higgs potential~\cite{Coleman:1973jx} is automatically finite.   Similar arguments can be applied to the gauge sector, as we show in App.~\ref{App:gauge}. In addition, both the coefficients of the quadratic $s_h^2$ and the quartic $s_h^4$ terms are at the same order in the Yukawa couplings implying that the model has the minimal universal fine-tuning needed to get a small $\xi = \sin^2(v/f)$, and the absence of double tuning.  The top Yukawa term is already $G_{V'}$ invariant so the top mass is also maximized, which suggests a relatively light lightest top partner,  further reducing the tuning  needed to obtain a 125 GeV Higgs. 

The paper is organized as follows. We first review the general formalism of pNGB's for symmetric spaces, and present the master formula for the low-energy effective Lagrangian in terms of a few form factors. We identify the maximal symmetry, which allows us to find the Coleman-Weinberg potential for general theories with such a maximal symmetry. We then apply our general results to  the minimal composite Higgs Model (MCHM)~\cite{Agashe:2004rs} based on the $SO(5)/SO(4)$ coset. We present a simple set of vector-like fermions, where the origin of the maximal symmetry can be nicely identified. Using collective symmetry breaking arguments we identify the form of the induced terms in the CW potential in agreement with the general result from the master formula. Next we discuss the tuning necessary to obtain a realistic Higgs sector, and explain why this model minimizes the tuning. Finally we show possible signals for maximal symmetry and conclusion. The Appendices contain a concrete realization of the maximal symmetry, more examples of the Higgs-parity operator for various cosets, the detailed symmetry structure of the gauge sector, the explicit expressions for the form factors of the MCH model, as well as the details of the numerical scan.

\section{Effective Lagrangian for pNGB's on Symmetric Spaces and Maximal Symmetry}
\label{sec:eff}

As usual in composite Higgs models, we will consider a strongly coupled system which dynamically breaks its global symmetry $G$ to $H$, and the Higgs fields are identified with the pNGBs which lie in the coset space $G/H$. The additional assumption we will make is that the coset space is a ``symmetric space", which means that it has the additional property that the commutator of two broken generators closes into the unbroken group $H$. While properties of such spaces have been studied before~\cite{symmetric}, the general formalism has not been commonly applied to composite Higgs models. First we summarize the basic features of symmetric spaces. The general structure of the commutation relations for the $T^{\hat{a}}(T^a)$ (un)broken generators is 
$[T^a, T^a] \sim T^a$, $[T^a, T^{\hat{a}}] \sim T^{\hat{a}}$ and $[T^{\hat{a}}, T^{\hat{a}}] \sim T^a$ where the first two relations are standard requirements such that $T^a$ form a subgroup, and the last relation is the added condition for the space to be symmetric. Some of the most commonly used moduli spaces satisfy this requirement, including $SU(N+M)/(SU(N)\times SU(M)\times U(1))$, $SO(N+M)/(SO(N)\times SO(M))$ and others. 
These conditions imply the existence of a parity operator $V$ (which is called Higgs-parity), which is an automorphism of the form $V T^a V^\dag  = T^a$ and $V T^{\hat{a}} V^\dag = - T^{\hat{a}}$~\footnote{This is the form of the automorphism for $SU(N+M)/(SU(N)\times SU(M)\times U(1))$,  $SO(N+M)/(SO(N)\times SO(M))$ or $SU(N)/SO(N)$ cosets. For the explicit form of $V$ for other coset spaces, see Appendix~\ref{App:Higgsparity}.}. As usual the pNGB fields $h^{\hat{a}}$ can be described by the Goldstone matrix 
\bea
U=\mbox{exp}\left(\frac{i h^{\hat{a}} T^{\hat{a}}}{f}  \right)\ .
\eea
The main consequence of the existence of the Higgs parity operator $V$  is that one can define a modified pNGB matrix which transforms linearly under the full set of symmetries $G$. The original pNGB matrix $U$ has the non-linear transformation properties~\cite{Coleman:1969sm,Callan:1969sn}
\bea
U\to g U h(h^{\hat{a}},g)^\dagger
\eea
where $g\in G$ and $h\in H$, and $h$ depends non-linearly on the pNGB field $h^{\hat{a}}$ and the transformation element $g$. However the parity transformed pNGB matrix $\tilde{U} = V U V = U^\dagger$ transforms as
 \bea
\tilde{U}\to VgV \tilde{U} h^\dagger\ .
\eea
We can then define the modified pNGB matrix $\Sigma'\equiv U \tilde{U}^\dagger V = U^2 V$, which transforms linearly under the full global symmetries
\bea
\Sigma' \to g \Sigma' g^\dagger \ .
\eea
The linearly realized global symmetry can be used to fully fix the structure of the low-energy effective Lagrangian of the theory. The SM fermions are charged under the $SU(2)_L\times U(1)_Y$ which is a subgroup of the full global symmetries $G$, thus they can always be embedded into the full symmetry group $G$. For the low-energy effective action we consider a spurionic embedding, which can always be written in the form $\Psi_{Q_L} =  \Lambda^\alpha_L Q_L ^\alpha$ and $\Psi_{t_R} =  \Lambda_R t_R$ if the  left-handed (LH) top doublet $Q$ and  right-handed (RH) top singlet $t_R$ are embedded into $\Psi_Q$ and $\Psi_{t_R}$, which are in some representation of the full global symmetry group $G$. Thus imposing the original $G$ symmetry will completely fix the most general effective action for the SM fermion fields coupled to the pseudo-Goldstone boson Higgses:
\bea 
\label{eq:Leff}
\mathcal{L}_{\mbox{eff}} &=& 
\bar{\Psi}_{Q_L} \slashed p(\Pi_0 ^q(p) + \Pi_1 ^q(p) \Sigma') \Psi_{Q_L} \nonumber \\
&+&   \bar{\Psi}_{t_R} \slashed p(\Pi_0 ^t(p) + \Pi_1 ^t(p) \Sigma') \Psi_{t_R} \nonumber \\
&+&   \bar{\Psi}_{Q_L}  M_1 ^t(p) \Sigma' \Psi_{t_R} +h.c.  
\eea
where the form factors $\Pi_{0,1} ^{q, t}$ and  $M_{1}^t$ encode the effect of the strong dynamics, and we assumed that $\Psi_{Q_L,t_R}$ are in the fundamental of $G$. In this case (since $\Sigma'^2=1$) only terms linear in $\Sigma'$ can show up. Using the spurions $\Lambda_{L,R}$ we can go back to the basis of the SM fermions to write the effective Lagrangian as 
\bea \label{eq:Lag_spurion}
\mathcal{L}_{ \mbox{eff} }&=& \bar{Q}_L ^\alpha \slashed p\mbox{Tr}[ (\Pi_0 ^q + \Pi_1 ^q \Sigma' ) P_{l}^{\alpha \beta} ] Q_L ^\beta 
\nonumber \\ &+& \bar{t}_{R} \slashed p \mbox{Tr}[(\Pi_0 ^t + \Pi_1 ^t \Sigma')P_r ]t_{R} \nonumber \\
&+& M_{1} ^t \bar{Q}_L ^\alpha t_R \mbox{Tr} [\Sigma'.P_{lr} ^\alpha]  \ ,
\eea
using the projection operators $P_{l,r,lr}$ defined from the spurions as $P_{l} ^{\alpha \beta} = (\Lambda_L ^{\beta })^{\dagger} \Lambda_L ^\alpha$, $P_{r} = (\Lambda_R )^{\dagger} \Lambda_R$ and $P^\alpha_{lr} = (\Lambda_R) ^{\dagger} \Lambda_L ^\alpha$.

A careful examination of the symmetries of the effective Lagrangian (\ref{eq:Leff}) will allow us to identify an enlarged global symmetry in certain limits, which we will call the maximal symmetry. This maximal symmetry is the key new ingredient of composite Higgs models which will be the focus of discussions for the rest of this paper. Let us first start with the massless limit of (\ref{eq:Leff})  when $M_1^t=0$ and also $\Pi_1^{q,t}=0$. In this case the global symmetry $G$ is enlarged to a chiral $G_L\times G_R$ symmetry acting on the left/right handed fermions  ${\Psi}_{Q_L}$/${\Psi}_{t_R}$. Now turning on the top mass term $\bar{\Psi}_{Q_L}  M_1 ^t(p) \Sigma' \Psi_{t_R}$ (while still keeping $\Pi_1^{q,t}=0$) we observe that we do not break the enlarged global symmetry completely, but rather leave a subgroup  $G_{V'}$ of $G_L \times G_R$ unbroken~\footnote{Another way to keep the Higgs potential finite is the left-right symmetric case $\Pi_0^q=\Pi_0^t$ and $\Pi_1^q=\Pi_1^t$, which is exactly the Weinberg sum rule chosen in Ref.~\cite{Marzocca:2012zn}. In this case, the top quark kinetic terms are invariant under the Higgs shift symmetry so only the mass term $M_1^t$ contributes to Higgs potential. We will present this case in a separate paper.}. We call this $G_{V'}$ the maximal symmetry, which is identified with the subgroup that keeps the pNGB field invariant $g_L \Sigma' g_R^\dag = \Sigma'$. We explain the structure of this maximal symmetry in more detail in App.~\ref{App:Maxsym}. The origin of this maximal symmetry and the conditions for the existence of this symmetry will be examined for the specific example of the $SO(5)/SO(4)$ minimal composite Higgs in the section below. One general property of the case with maximal symmetry is that the Higgs potential is simply given by the top mass square  (up to some form factor coefficients from the spurion matrix and the momentum integration)
\bea 
\label{eq:Higgspotential}
V(h) =-2N_c \int \frac{d^4 p}{2\pi^4} \mbox{log}\left[1+\frac{ (M^t_1)^2 |\mbox{Tr} [\Sigma'.P_{lr} ^1 ]|^2 }{ p^2 \mbox{Tr}[ \Pi_{0} ^q P_{l} ^{11}] \mbox{Tr}[ \Pi_{0} ^t P_{r}]} \right].
\eea  
The numerator $(M^t_1)^2 |\mbox{Tr} [\Sigma'.P_{lr} ^1 ]|^2$ in the above expression~(\ref{eq:Higgspotential}) determines all the properties of the Higgs potential, and as we will see later will imply that the Higgs potential is actually finite.

\section{$SO(5)/SO(4)$ MCHM}
\label{sec:so5}

We have seen the potential emergence of maximal symmetry in composite Higgs models from the analysis of the effective Lagrangian.  Here we present an explicit realization of a realistic model with the maximal symmetry. We show the conditions for the emergence of maximal symmetry, as well as its consequences on the UV properties of the Higgs potential. We will see that the existence of maximal symmetry will impose a condition on the spectrum of the composites as well as relations among the mixing terms between the elementary and the composite sectors.  The MCHM is based on the smallest $SO(5)/SO(4)$ coset space with custodial symmetry~\cite{Chang:2003un,Agashe:2004rs}, so we will choose this as our benchmark example. This coset does correspond to a symmetric space, thus the general formalism presented above can be applied here. The Goldstone matrix for the fields $h^{\hat{a}}$ corresponding to the broken generators is
\bea
U=\mbox{exp}\left(i \frac{\sqrt{2}}{f} h^{\hat{a}} T^{\hat{a}}  \right) \ ,
\eea
which transforms non-linearly as $U \to g U h(h^{\hat{a}},g)^\dagger$, $g\in SO(5)$ and $h\in SO(4)$. The explicit form of the Higgs parity operator $V$ is 
\bea
V=\left( \begin{array}{cc}
{ \bf 1}_{4\times 4} & 0  \\ 
0 & -1  \\ 
\end{array}  \right)
\eea 
with the properties $V=V^\dagger$ and $V^2=VV^\dagger =1$. As explained above we can then construct the linear Goldstone matrix $\Sigma' =U^2 V$, which is the variable that should show up in the low-energy effective Lagrangian. 

Next we will explicitly construct this low-energy effective Lagrangian for the fermion sector obtained from the interactions with the heavy top-partners. This will also allow us to explain the origin and the significance of the maximal symmetry. Following the usual assumption of partial compositeness, the SM fermions, and in particular the third generation quarks $q_L =(t_L ,\, b_L) $ and $t_R$ are introduced to couple linearly to the strong sector. Thus we assume that the elementary-composite interaction is~\cite{Matsedonskyi:2012ym}
\bea \label{eq:SO5mixing}
\mathcal{L} = \lambda_L \bar{q}_L ^\alpha \Lambda_{\alpha I} ^L \mathcal{O}_R ^I +\lambda_R \bar{t}_R  \Lambda_{I} ^R \mathcal{O}_L ^I  +h.c 
\eea   
where $\mathcal{O}_{L,R}$ are fermionic operators from the composite sector. These $\mathcal{O}_{L,R}$ transform in a linear representation of $SO(5)$, $\alpha$ is an $SU(2)$ index and $I$ is an $SO(5)$ index. The $\Lambda^{L,R}$ are spurions characterizing the nature of the explicit breaking arising from the fermion sector.  The mixing terms will be $SU(2)_L \times U(1)_Y \times SO(5)$ invariant if the spurions $\Lambda^{L,R}$ transform as  
\bea
\Lambda^{L,R} \to  u \Lambda^{L,R} g^\dagger 
\eea  
where $u$ is an electroweak transformation and $g$ is a global $SO(5)$ transformation.  The actual values of the spurions $\Lambda ^{L,R}$ are uniquely fixed by the requirement of leaving the SM $SU(2)_L$ subgroup embeded in $SO(5)$ unbroken~\cite{Matsedonskyi:2012ym}: 
\bea
\Lambda ^{L} = \frac{1}{\sqrt{2}} \left( \begin{array}{ccccc}
0 & 0 & 1 &-i & 0 \\ 
1 & i & 0&0 &0
\end{array}  \right),   \Lambda ^{R} = \left( \begin{array}{ccccc}
0 & 0 & 0& 0& 1 
\end{array}  \right) .
\eea 
Another way to state this is that the transformation properties of the spurions $\Lambda^{L,R}$ will fix how to embed the SM fermions into incomplete $SO(5)$ multiplets. In this approach we will have  $q_L$ and of $t_R$ embeded into the ${\bf 5}$ of SO(5) (together with a proper $U(1)_X$ charge assignment):
\bea
\Psi_{q_L} =\frac{1}{\sqrt{2}} \left( \begin{array}{c}
b_L \\ 
-i b_L \\ 
t_L \\ 
i t_L \\ 
0
\end{array} \right) \quad  \Psi_{t_R} = \left( \begin{array}{c}
0 \\ 
0 \\ 
0 \\ 
0\\ 
t_R
\end{array} \right)
\eea
The $\mathcal{O}$ composite fermions are assumed to be Dirac fermions, with Dirac masses arising for each of them from the composite dynamics. 
The operators $\mathcal{O}_{L,R}$ will be contained in some of the $SO(5)$ representations, for example ${\bf 1}$, ${\bf 5}$, ${\bf 10}$, or  ${\bf 14}$. Here we will consider the case where $\mathcal{O}$ is contained in the ${\bf 5}$ of $SO(5)$, but our analysis can be directly generalized to other representations.  The decomposition of ${\cal O}$ under $SO(4)$ is ${\bf 5} \to {\bf 4}+ {\bf 1}$, or $\mathcal{O} \to  \Psi_Q +\Psi_S$, where  $\Psi_{Q,S}$ contain the top partners~\cite{DeSimone:2012fs} 
\bea
\Psi_Q = \frac{1}{\sqrt{2}}\left( \begin{array}{c}
i B -i X_{5/3} \\ 
B + X_{5/3} \\ 
i T + i X_{2/3} \\ 
-T +X_{2/3} \\
0
\end{array} \right) \quad \Psi_S =\left(  \begin{array}{c}
0 \\ 
0 \\ 
0 \\ 
0 \\ 
T_1
\end{array} \right)  
\eea
The general fermionic Lagrangian (\ref{eq:SO5mixing}) can then be parametrized as~\cite{Marzocca:2012zn} 
\bea \label{eq:SO5mixing1} 
\mathcal{L}_{f} &=& \bar{\Psi}_Q (i \slashed \bigtriangledown -M_Q ) \Psi_Q + \bar{\Psi}_S (i \slashed \bigtriangledown -M_S ) \Psi_S \nonumber \\ 
&+& \frac{f }{\sqrt{2}}\bar{\Psi }_{t_R }P_L( \epsilon_{tS}   U \Psi_{S} + \epsilon_{tQ} U \Psi_{Q}  ) \nonumber \\
&+& f \bar{\Psi }_{q_L} P_R (\epsilon_{qS}  U \Psi_S + \epsilon_{qQ}  U \Psi_Q)  +h.c.,            
\eea
where $\lambda_{L,R}$ are contained in the definitions of the Yukawa couplings $\epsilon$, and top and top partner masses are 
\bea \label{eq:mtop}
m_t = \frac{\epsilon_{qQ} \epsilon_{tS} f^2 }{2 M_T M_{T_1}} \left|\frac{\epsilon_{qS} } {\epsilon_{qQ} } M_Q - \frac{\epsilon_{tQ} } {\epsilon_{tS} } M_S \right| \sin \frac{\langle h \rangle}{f}
\eea  
\bea
M_T = \sqrt{\epsilon_{qQ} ^2 f^2 +M_Q ^2}, \ 
M_{T_1} = \sqrt{\frac{\epsilon_{tS} ^2}{2} f^2  +M_S^2}.
\eea

In order to understand the symmetry properties of this Lagrangian more easily it is useful to combine $\Psi_Q$ and $\Psi_S$ back to complete representations  ${\bf 5}$ of the global symmetry $SO(5)$ (and assume for simplicity that CP is conserved):   
\bea
\Psi_+  = \frac{1}{\sqrt{2}} (\Psi_Q +\Psi_S ) \quad \Psi_-  = \frac{1}{\sqrt{2}} (\Psi_Q -\Psi_S )   \ .
\eea       
Thus $\Psi_{+}  $ and $\Psi_{-} $ are related by the Higgs parity operator: $\Psi_+  =V \Psi_-$, and are not  independent fields. 
 
Our original fermion Lagrangian (\ref{eq:SO5mixing1}) in terms of $\Psi_\pm$ is:
\bea
\label{eq:fulleq}
&\mathcal{L}_{f}& =  2 \bar{\Psi}_+   i \slashed \bigtriangledown   \Psi_+  
 + f( c_{-R} \bar{\Psi }_{t_R } U V \Psi_{+L}  + c_{+R} \bar{\Psi }_{t_R}  U \Psi_{+L}  )   \nonumber \\
 &-&(M_Q +M_S) \bar{\Psi}_{+L} \Psi_{+R} -(M_Q -M_S) \bar{\Psi}_{+L}  V\Psi_{+R}  
       \nonumber \\      
&+& f( c_{-L} \bar{\Psi }_{q_L} U V \Psi_{+R}  + c_{+L} \bar{\Psi }_{q_L} U  \Psi_{+R} ) 
+h.c.,
\eea 
where the Yukawas are $c_{\pm R} =\frac{\epsilon_{tQ} \pm  \epsilon_{tS}  }{2}$, $c_{\pm L} =\frac{\epsilon_{qQ} \pm \epsilon_{qS} }{\sqrt{2}}$. 

This simple form of the Lagrangian allows us to identify the possible symmetry breaking patterns and identify the conditions for the emergence of the maximal symmetry. We have assumed here that the composite fermions $\Psi_Q$ and $\Psi_S$ fill out a full $SO(5)$ representation. This does not generically have to be the case, but it will be a necessary condition on the spectrum of composites in order to obtain maximal symmetry. Once the composites do fill out a complete $SO(5)$ representation the kinetic terms will have the enlarged $SO(5)_L \times SO(5)_R$ chiral global flavor symmetry, which can have various symmetry breaking patterns depending on the structure of the Yukawa couplings and composite mass terms. These symmetry breaking patterns will determine the form of the radiatively induced Higgs potential and its degree of divergence.
Since our goal is to find an implementation of the maximal symmetry, we will set $c_{-L} = c_{-R}=0$ in the general Lagrangian. If $c_-$ and $c_+$ were to appear simultaneously in the Lagrangian one would not be able to maintain an entire $SO(5)$ global symmetry as needed for maximal symmetry. This requirement for maximal symmetry is equivalent to the assumption that the elementary-composite mixing terms are fully $SO(5)$ invariant.  Of course one could as well have chosen $c_{+L,R}=0$ and arrive at similar results. In this case, the Lagrangian is
\bea
\label{eq:mainL}    
& \mathcal{L}_{f}& = 2 \bar{\Psi}_+  i \slashed \bigtriangledown   \Psi_+  + f c_{+R} \bar{\Psi }_{t_R} U \Psi_{+L} + f c_{+L} \bar{\Psi }_{q_L}  U  \Psi_{+R}  \nonumber \\  &-&  \bar{\Psi}_{+L} ((M_Q +M_S) +(M_Q -M_S)  V)\Psi_{+R} + h.c.
\eea 

Once we impose $c_{-L,R}=0$ the mixing terms will have the full $SO(5)_L \times SO(5)_R$ chiral global symmetry, and the breaking pattern depends on the relation of the mass terms $M_{Q,S}$, giving rise to the following possible breaking patterns:
\bea
M_Q- M_S = 0 & \Rightarrow & SO(5)_L \times SO(5)_R /SO(5)_{V} \nonumber \\
M_Q +M_S = 0 & \Rightarrow & SO(5)_L \times SO(5)_R /SO(5)_{V'} \nonumber \\
|M_Q| \neq  |M_S| & \Rightarrow & SO(5)_L \times SO(5)_R /SO(4)_{V} 
\eea
Clearly the second case $M_Q+M_S=0$ corresponds to the maximally symmetric scenario, which we will eventually be focusing on. Let us now examine what these symmetries imply for the structure of the Higgs potential.

\begin{itemize}
\item If $M_Q=M_S$ the second (twisted) mass term vanishes. The entire remaining Lagrangian is invariant under the $SO(5)_V$ global symmetry where $U\Psi_{+L,R} \to {\cal V} U \Psi_{+L,R}, \Psi_{t_R,Q_L}\to {\cal V} \Psi_{t_R,Q_L}$. This global symmetry contains the original shift symmetry, so the entire Higgs potential vanishes, thus every term must be proportional to $M_Q-M_S$.
\item If the untwisted mass vanishes $M_Q+M_S=0$, then there is still a remaining global symmetry, the maximal symmetry $SO(5)_{V'}$, but it does {\it not} contain the entire Goldstone shift symmetry, thus a potential will be generated. The transformation here is $U\Psi_{+L} \to L U\Psi_{+L}, U\Psi_{+R}\to R U\Psi_{+R}$. Since the twisted mass term can be also written as $\bar{\Psi}_{+L} U^\dagger \Sigma' U \Psi_{+R}$, the condition for the unbroken $SO(5)_{V'}$ symmetry is $L^\dagger \Sigma' R =\Sigma'$. 
\end{itemize}
In order to find the actual structure of the radiatively induced Higgs potential we need to examine the collective symmetry breaking properties of (\ref{eq:mainL}). 
\begin{itemize}
\item The combination of the $c_{+L}$ and the two mass terms will break the shift symmetry. However we can see that we need all three of these terms to generate a potential. If $c_{+L}=0$ we don't have $U$ appearing at all. If $M_Q-M_S=0$ we have the vectorlike $SO(5)_V$ symmetry as above. If $M_Q+M_S=0$ we have the unbroken global symmetry $U\Psi_{+R} \to R U\Psi_{+R}$ and $\Psi_{+L}\to VU^\dagger R U \Psi_{+L}$ which contains the Higgs shift symmetry. Thus the Higgs potential must be proportional to $c_{+L} (M_Q+M_S)(M_Q-M_S)$, and the left-handed top $\Psi_{q_L}$ in the closed loop contributed to the Higgs potential can only couples through $f c_{+L} \bar{\Psi }_{q_L}  U  \Psi_{+R} + h.c.$ so $c_{+L}$ actually has to show up as $|c_{+L}|^2$, resulting in a contribution logarithmically sensitive to the cutoff: 
 \bea
V_{L\xi} \sim |c_{+L}|^2 f^2 (M_Q+M_S)(M_Q -M_S) \log \Lambda^2 
\eea    
A similar term is obtained using $c_{+R}$: 
\bea
V_{R\xi} \sim  |c_{+R}|^2 f^2 (M_Q+M_S)(M_Q -M_S) \log \Lambda^2 
\eea 

\item The combination of $c_{+L},c_{+R}$ and the twisted mass term will break the shift symmetry (but leave the maximal symmetry intact), and a potenial will be generated. Again we can see we need all three terms to generate a potential. If the twisted mass term is turned off we again have the vectorlike $SO(5)_V$ containing the shift symmetry. If for example $c_{+L}$ is turned off, we again have the global symmetry $U\Psi_{+R} \to R U\Psi_{+R}$ and $\Psi_{+L}\to VU^\dagger R U \Psi_{+L}$ which contains the Higgs shift symmetry. So we need all three terms to show up, and in fact to be able to actaully generate a potential all three have to show up twice, giving rise to a 
finite contribution of the form. 
\bea
|c_{+L}|^2 |c_{+R}|^2 f^4 (M_Q -M_S)^2  / \Lambda^2.
\eea     

\comment{
\begin{figure}
\begin{center}
\includegraphics[width=5cm]
{fig/aa.eps}
\end{center}
\caption{$1-$loop contribution of the SM top and top partners to Higgs potential through Yukawa couplings. A grey blob denotes the Yukawa couplings $fc_{+L} $ or $fc_{+R}$ and the circle cross denotes the mass of top partners.   } 
\label{fig:feyn}  
\end{figure} 
}
     
\end{itemize} 

Integrating out the heavy top partner $\Psi_+$ from the Lagrangian in (\ref{eq:fulleq}) we obtain the form factors $\Pi_{0} ^{q,t}$, $\Pi_{1} ^{q,t}$ and $M_1 ^t$ for the effective Lagrangian of the elementary quarks as in Eq.~(\ref{eq:Leff}).  The explicit expressions of the form factors are given in App.~\ref{App:formfactors}. Recalling that the effect of the $SO(5)_L\times SO(5)_R$ global symmetry on the elementary fields is $\Psi_{t_R}\to R \Psi_{t_R}, \Psi_{Q_L}\to L \Psi_{Q_L}$, it is clear that $\Pi_0 ^{q}$($\Pi_0 ^{t}$)  is $SO(5)_{L}$($SO(5)_R$) invariant, while $\Pi_{1} ^q$ ($\Pi_{1} ^t$) break the full $SO(5)_L\times SO(5)_R$ to $SO(5)_V$ coresponding to $L=R$. However the top mass term $M_1^t$ leaves the maximal $SO(5)_{V'}$ invariant, since that symmetry corresponds to the choice where $L^\dagger \Sigma' R =\Sigma'$. Thus for the maximally symmetric $SO(5)_{V'}$ case we automatically get $\Pi_{1} ^{q,t}=0$. The expression for the top mass in this case simplifies to $m_t = c_{+L} c_{+R} (M_Q-M_S) f^2/(2M_TM_{T_1})$, and we see that the contribution to the Higgs potential is proportional to the top mass square $V \sim (M_1^t \Sigma')^2 \sim \lambda_L^2 \lambda_R^2 f^4 (M_Q-M_S)^2 / \Lambda^2$, which is finite and has the form as expected from the general symmetry arguments. 

We summarize this section by restating the conditions for maximal symmetry: the composites should fill out a full $SO(5)$ representation, the elementary-composite mixing terms should be fully $SO(5)$ invariant, then the twisted mass term for the composites preserves the maximal symmetry $SO(5)_{V'}$ (while the untwisted mass term should vanish).

\section{Tuning in the Higgs Potential}

Parametrizing the potential as usual as 
\bea
V(h) = -\gamma s_h^2 +\beta s_h^4 
\eea
we find at the minimum $\xi \equiv s_h^2 = \frac{\gamma}{2\beta}$. Our main result on tuning is that as a result of maximal symmetry the fermionic contribution to $\gamma$ and $\beta$ are equal. Hence the only source of tuning is the approximate cancellation between the fermionic and gauge contributions to $\gamma$ implying $\gamma_f +\gamma_g \approx 0$, yielding in a minimally tuned composite Higgs model. Below we present a detailed explanation of this result.

In generic composite Higgs models the Higgs potential is usually (quadratically) divergent, with 
\bea
 \gamma_f &=& \frac{N_c M_f^4 }{16 \pi^2 g_f ^2} \left[ c_2 \epsilon ^2 \frac{\Lambda^2}{M_f^2} 
+c_0 \epsilon^4 \log \frac{\Lambda^2}{M_f^2} + {\rm finite} \right] \nonumber \\
 \beta_f &=& \frac{N_c M_f^4 }{16 \pi^2 g_f ^2} \left[  
c'_0 \epsilon^4 \log \frac{\Lambda^2}{M_f^2} + {\rm finite} \right]
\eea
where $\epsilon$ is a Yukawa coupling $ \epsilon_{qS(Q)}, \epsilon_{tS(Q)}$ and $M_f$ is a typical fermion resonance mass with interaction strength $g_f$, and $M_f = g_f f$.  To obtain $\xi \ll 1$ requires that we first tune the quadratically divergent coefficient $c_2$ (by cancelling various ${\cal O} (\epsilon^2)$ contributions against each other) such that the quadratically divergent contribution gets reduced to the size of the log divergent term. This impies a tuning of the order of the ratio of the two contributions to $\gamma$. In addition, one needs to ensure that $\gamma = 2 \xi \beta$, which implies another tuning of order $1/\xi$.  The total tuning will be of the  order
\bea
\Delta \simeq  \frac{1}{\xi} \frac{g_f ^2  }{ \epsilon^2} \frac{\Lambda^2 }{M_f^2 \mbox{log} \frac{\Lambda^2}{M_f ^2}}  
\eea 
which is paremetrically much larger than the minimal tuning $\Delta_{\mbox{min}} = {1}/{\xi}$~\cite{Panico:2012uw}. 
 
Holographic composite Higgs~\cite{Contino:2003ve} models based on a warped extra dimension and their deconstructed versions~ \cite{Cheng:2006ht,Foadi:2010bu,Panico:2011pw,DeCurtis:2011yx} yield a log divergent or finite Higgs potential. For symmetric spaces the low-energy effective Lagrangian after integrating out the heavy fermions is still given by~(\ref{eq:Leff}), except the form factors will be more strongly suppressed at large momenta due to the additional fermion propagators needed to be inserted for the additional intermediate sites (in deconstructed versions) or propagation in the bulk (in extra dimensional versions). For example for a three site model, the leading local contribution to the Higgs potential arises from dimension six operators, implying that $\Pi_{1} ^{q,t}$ behaves as  $\mathcal{O}(p^{-6})$ for large $p$ and thus $\gamma$ and $\beta$ are finite (See Appendix~\ref{App:deconstruction} for details). However  $\gamma$ is still  $\mathcal{O}(\epsilon^2)$ while $\beta$ is $\mathcal{O}( \epsilon^4)$ in the Yukawa insertions. Thus the discrete MCHM$_5$ has a double tuning given by $\Delta^{5+5} \simeq \frac{1}{\xi} \frac{g_f ^2}{\epsilon^2}$ ~\cite{Panico:2012uw}, which is bigger than the minimal tuning for $\epsilon < g_f$.

However the model with maximal symmetry presented in (\ref{eq:mainL}) does not suffer from this double tuning, but rather has the minimal tuning $1/\xi$. Thus maximal symmetry implies minimal tuning. A simple way to see this is to realize that for models with maximal symmetry the Higgs potential will have an additional $Z_2$ symmetry corresponding to the $s_h \to -c_h$ exchange, analogous to the case of twin higgs models (where the exchange symmetry is a consequence of the $Z_2$ symmetry between the visible and twin sectors). Here instead one has another $Z_2$ symmetry of the form:
\bea
\Psi_{+L} &\to & P_1 \Psi_{+L}, \  \Psi_{+R} \to V P_1 V \Psi_{+R}, \ U\to V U VP_1V,  \nonumber \\
\Psi_{q_L} &\to &  V \Psi_{q_L}= \Psi_{q_L}, \   \Psi_{t_R} \to P_2 \Psi_{t_R}=\Psi_{t_R}       
\eea      
where $P_1 ={\rm diag} (1_{3\times 3}, \sigma_1 ), P_2 = {\rm diag} (1_{3\times 3}, -\sigma_3)$. Using $VP_1V=P_1$ one can easily show that this leaves (\ref{eq:mainL}) invariant,  while the effect 
of this transformation on the Goldstone matrix $U$ is the exchange $s_h \Leftrightarrow -c_h$, implying that the Higgs potential must be invariant under this exchange symmetry. This symmetry will then forbid the $\epsilon^2 s_h^2$ term (similar to twin composite Higgs models~\cite{Geller:2014kta,Barbieri:2015lqa,Low:2015nqa}) and eliminate the double tuning.\footnote{Another solution for eliminating double tuning within composite Higgs models was presented in~\cite{Javi}, where an embedding of $t_L$ into a ${\bf 14}$ of $SO(5)$ was used to achieve this.} 

The explicit expression of the Higgs potential in our model with maximal symmetry  up to $\mathcal{O} ({c_{+L} ^2 c_{+R} ^2}/{g_f ^4})$ using~(\ref{eq:Higgspotential}) will be
\bea
V_{h} &\simeq & c_{LR} \frac{N_c M^2_{f} (M_S-M_Q)^2 }{16\pi^2 g_f^4 } \left( \frac{c_{+L} ^2 c_{+R}^2}{g_f^4} \right)    [-s_h ^2 +s_h ^4 ]   \nonumber \\ 
 &\simeq & c_{LR} \frac{N_c M^4_{f} }{16\pi^2  } \left( \frac{y_t}{g_f} \right)^2   [-s_h ^2 +s_h ^4 ]  
\eea
where $y_t = {m_t}/{v}  \simeq c_{+L} c_{+R}/{g_f} $ is the top Yukawa coupling and $c_{LR}$ is an order one dimensionless constant. Thus the leading fermion loops result in $\beta = \gamma$, and an almost constant vacuum alignment parameter $\xi \simeq 0.5$. In order to reduce $\xi$ to experimentally allowed values $\xi \ll 1$ one needs to include gauge contributions,  and 
impose a cancelation between the fermionic and gauge contributions of the $\gamma$ terms $\gamma_f \simeq -\gamma_g $ (while $\beta_g $ is at order $\mathcal{O}(g^4/g_\rho ^4 )$ which is always negligible compared to $\beta_f$). The tuning required will then be
\bea
\Delta^{(\bf 5 + 5 )} = \frac{\mbox{max}(|\gamma_f |, |\gamma_g|)}{|\gamma_f +\gamma_g | } \simeq \frac{1}{2\xi}
\eea
which is the minimal universal tuning necessary for a small $\xi$. As discussed in App.~\ref{App:gauge}, imposing that the vector meson $\rho_\mu$ and the axial-vector meson $a_\mu$ form a full adjoint representation of $SO(5)$ automatically renders the higgs potential finite, and   the corresponding gauge contributions to the potential are~\cite{Marzocca:2012zn}  
\bea
\hspace*{-0.5cm}\gamma_g =- \frac{9f^2g^2m_\rho ^2 \ln 2}{64\pi^2}, ~ \beta_g = \frac{9f^4 g^4}{1024\pi^2}\left(5+\mbox{log} \frac{m_W ^2}{32 m_\rho ^2} \right)  .
\eea

For general composite Higgs models one usually needs some additional tuning to get the Higgs mass down to 125 GeV. However the model with maximal symmetry has the special property 
that the top mass is maximized:  $m_t \sim \sin \theta_L  \sin \theta_R$ $|M_Q-M_S|s_h$ where 
$\theta_L$ and $\theta_R$ are the degrees of LH and RH top compositeness. Since maximal symmetry implies $M_Q=-M_S$, the $|M_Q-M_S|$ factor is maximized, hence the degree of compositeness can be minimized while the top mass is held fixed at the physical value. This also implies that the mass of the lightest top partner $\mbox{min}\{M_T,M_{T_1}\}=\mbox{min}\{\frac{M_S}{\cos \theta_L},\frac{M_{Q}}{\cos \theta_R} \}$ is also automatically reduced, which in turn cuts off the top contribution to the Higgs mass earlier,  $m_H \varpropto {\mbox{min}\{M_T,M_{T_1}\}} m_t /{f}$, and allows us to obtain a light 125 GeV Higgs in the maximally symmetric limit.  

To explicitly verify our estimates we have numerically evaluated the tuning in the model of (\ref{eq:mainL}). We have used the measure~\cite{Panico:2012uw} of tuning
\bea
\Delta_m =\mbox{max}(\Delta_i), \quad \Delta_i=\left|\frac{2x_i}{s_h} \frac{c_h^2}{f^2 m_h ^2} \frac{\partial^2 V_h}{\partial x_i \partial s_h}  \right|
\eea
where the $x_i$'s are the parameters of the theory. The maximal symmetry implies   
$M_S =- M_Q$ and $\epsilon_{tQ} = \epsilon_{tS}=c_{+R}, \epsilon_{qQ} = \epsilon_{qS}=\frac{c_{+L}}{\sqrt{2}}$, which we have imposed here. The analytical expression for the maximal tuning (using the above measure) at order $\mathcal{O} (y_t^2 / g_f^2)$ is 
\bea
\label{eq:analytunning}
\Delta_m \simeq 2\gamma_g /|\gamma_f +\gamma_g | \simeq \frac{1}{\xi} - 2 \ .
\eea
 
\begin{figure}
\begin{center}
\includegraphics[width=8.7cm]
{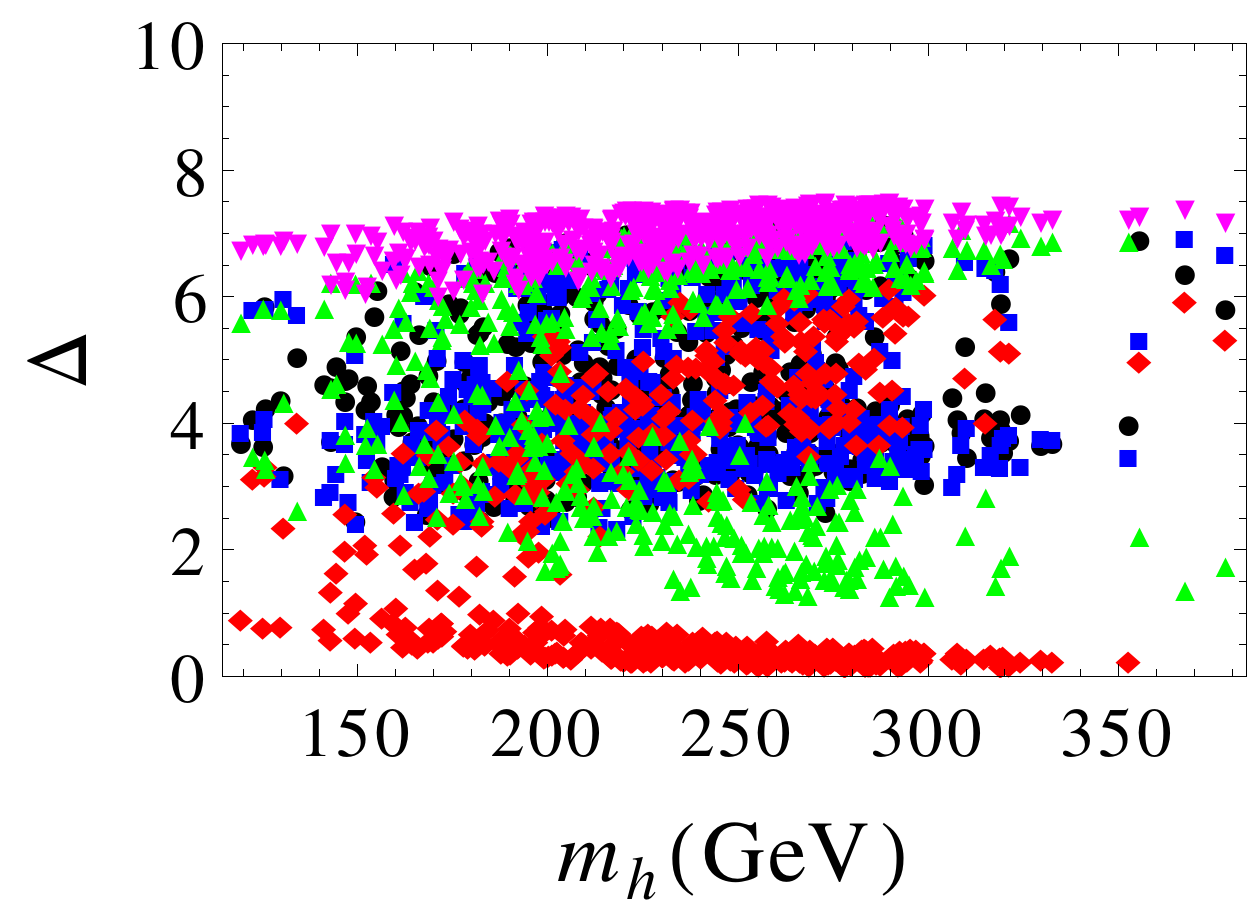}

\includegraphics[width=9cm]
{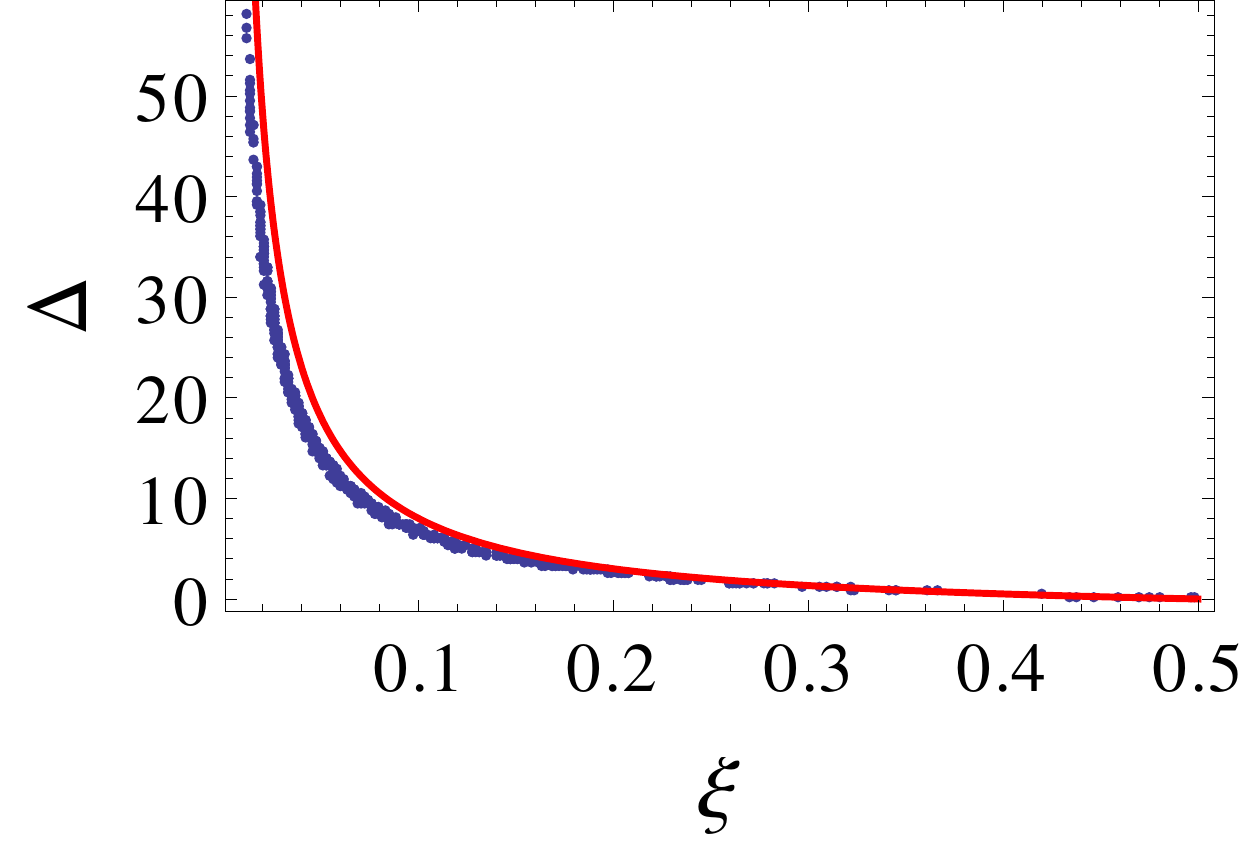}
\end{center}
\caption{Top: Scatter plot of tuning $\Delta_i$ for the various input parameters $x_i$, $c_{+R}$ (black), $c_{+L}$ (blue), $f$ (red), $M_S$ (green) and $m_\rho $ (magenta), as a function of $m_h$ with $\xi =0.1$ held fixed. Bottom: the tuning $\Delta_m$ as a function of $\xi$ for higgs mass $m_h =125$ GeV. The red solid line is the analytic result from Eq.~(\ref{eq:analytunning}).} 
\label{fig:tuning}  
\end{figure}

The numerical values of the tuning are shown in the top panel of Fig.~\ref{fig:tuning} as a scatter plot for the contribution from the different fundamental parameters $x_i$, which are chosen to be $c_{+R}$(black), $c_{+L}$(blue), $f$(red), $M_S$(green), and $m_\rho $(magenta), as a function of $m_h$ with $\xi =0.1$ held fixed. In the bottom panel the amount of tuning as a function of the vacuum alignment parameter $\xi$ are shown $m_h =125$ GeV. We can clearly see that the largest tuning is from $m_\rho$ which is from the requirement $\gamma_f \simeq - \gamma_g$ and is slightly smaller than 8 for $\xi=0.1$ because corrections beyond those at $\mathcal{O} (c_{+L}^2 c_{+R}^2 / g_f^4)$ can also contribute to $\beta_f$ 
making it slightly smaller than $\gamma_f$. 
     
\section{Signals of Maximal Symmetry}

The main consequence of maximal symmetry on the general effective Lagrangian in Eq.~(\ref{eq:Leff}) is the vanishing of the form factors $\Pi_1^{q,t} = 0$, which is what one would like to check experimentally. The best way to do that is to consider the properties of the top quark. In Eq.~(\ref{eq:Leff}) one can canonically normalize the top quark field, such that the form factors
now appear in the top mass term:
\bea  \label{eq:Lag_nonlinear}
\mathcal{L}_{\mbox{eff}} =   \frac{ M_1 ^t \mbox{Tr} [\Sigma'.P_{lr} ^1] \bar{t}  t}{\sqrt{ \mbox{Tr}[ (\Pi_0 ^q + \Pi_1 ^q \Sigma') P_{l}^{1 1} ]  \mbox{Tr}[ (\Pi_0 ^t + \Pi_1 ^t \Sigma') P_{r}]}    }   
\eea  
Expanding this in terms in powers of the Higgs field will give the various top-top-Higgs couplings, which (at least $tth$ and $tthh$) should be measurable at the LHC. 
The presence of the $\Pi_1 ^{q,t}$ form factors in (\ref{eq:Lag_nonlinear}) make the top Yukawa to depend on more than a single trigonometric function of $h/f$, which is absent for the case of maximal symmetry. Thus precise measurements of the top-Higgs couplings could be used to test maximal symmetry. For example, in MCHM$_5$, the top Yukawa couplings can be parametrized as 
\bea
\mathcal{L}_Y \sim M_1 ^t \sin \frac{2h }{f}  \left( 1 + (\alpha_q   \Pi_1 ^q +  \alpha_t  \Pi_1 ^t)\sin^2 \frac{h }{2f} \right) \bar{t} t.
\eea
The prescription would be to first measure the top higgs coupling in ordere to guess the form of $\mbox{Tr} [\Sigma'.P_{lr} ^1]$ (which also depends on the representation of the fermions). A more precise measurement of those couplings at future colliders can then tell us the whether the top mass term can be written in terms of a single trigonometric function or not.

Another way to test maximal symmetry is via the properties of the additional resonances if they are within the reach of the LHC (or future colliders).  One can then derive sum rules for the conditions of the cancellation of the quadratic and log divergences in the Higgs potential. For example for the  case of the top partners, we obtain the sum rules~\footnote{We thank Tao Liu for sharing a preliminary version of~\cite{TN} with us  containing~(\ref{eq:sumrule1}).}
\bea 
\label{eq:naturalness}
\mbox{Tr}[Y_m M_D] &=& 0 +\mathcal{O}(v^2/M_f^2)    \label{eq:sumrule1} \\
\mbox{Tr}[Y_m M_D^3]&=& 0 +\mathcal{O}(v^2/M_f^2)  \label{eq:sumrule2}
\eea
where $Y_m$ is the Yukawa coupling matrix of the top partners and and the top quark while $M_D$ is their mass matrix. The first (second) condition is the cancellation of quadratic (log) UV divergences in the Higgs mass term. The derivation of the above formulae for the general case (including scalars) will be presented elsewhere. Measuring the masses and couplings of all charge $2/3$ top partner resonances, one can test these sum rules and thereby maximal symmetry.

\section{Discussions and Conclusions} 

In this letter, we explored models of radiative EWSB where the Higgs is a pNGB of a symmetric coset space $G/H$. In this case, there exists an unique Higgs parity operator $V$, and a modified pNGB matrix $\Sigma'$ can be constructed which transforms linearly under the full global symmetries. This symmetry fixes the structure of the general low-energy effective Lagrangian between the SM fields and the GB matrix to generate the effective Higgs potential. We applied our results to study the top-Higgs system, and found that there might be an enhanced global symmetry (which we call maximal symmetry) $G_{V'}$ which is the maximal subgroup of the chiral symmetry $G_L \times G_R$ for  LH and RH top quarks. This maximal symmetry implies that the Higgs potential is automatically UV finite, and the tuning of the Higgs potential is also minimized. The origin of the minimal tuning is that the quadratic term from the top sector is suppressed, while the physical Higgs mass is automatically small due to the maximized top mass term. We have applied this maximal symmetry to MCHM$_5$ where only one free parameter is allowed for a given $\xi$ and confirmed numerically that even in the simplest case, our model has minimal universal tuning $1/2\xi$. Testing our model requires either accurate measurements of the top-multi-Higgs couplings or testing the sum rules for the masses and couplings of the heavy resonances implied by the cancellation of the divergences in the Higgs mass term. 

\section{Acknowledgements} 

We thank Brando Bellazzini, Tao Liu, Michael Geller and Javi Serra for useful discussions and comments. C.C. thanks the KITPC in Beijing for its hospitality while this project was initiated, and the Aspen Center for Physics while this this work was in progress. C.C. is supported in part by the NSF grant PHY-1316222. J.S. is supported in part by the Project 11647601, 11675243 supported by NSFC. T.M. is supported in part by project Y6Y2581B11 supported by 2016 National Postdoctoral Program for Innovative Talents.

\appendix

\section{A Concrete Realization of the Maximal Symmetry}
\label{App:Maxsym}

We present an illustration of the appearance of the maximal symmetry. 
Consider a general $G_L \times G_R$ chiral symmetry broken to $G_{V'}$ through a twisted link field ${\Sigma}'$, where (up to a $G$ transformation) this twisted link field ${\Sigma}'$ has a twisted VEV $V$, and also serves as the automorphism map for the symmetric space $G/H$.  As usual $V$ has the properties  $VT^{a}V^\dag=T^{a}$ for the unbroken and  $VT^{\hat{a}}V^\dag =-T^{\hat{a}}$ for the broken generators.   The unbroken group is given by $LVR^\dag = V$, or $LV = VR$. To find the actual unbroken combination of generators we take the explicit forms $L=\textrm{exp}(i \theta_L^a T_L^a)$ and $R=\textrm{exp}(i \theta_R^b T_R^b)$. Considering infinitesimal transformations we get  $\theta_L^a T_L^a V = V \theta_R^b T_R^b$. Since $V$ is the Higgs parity operator: $V T^{a} = T^{a} V$ and $V T^{\hat{a}} =-T^{\hat{a}} V$, we find that the unbroken $G_{V'}$ symmetry contains the combination of generators $\theta_L^a = \theta_R^a$ for the unbroken direction and $\theta_L^{\hat{a}} = - \theta_R^{\hat{a}}$ for the broken direction. Therefore, the twisted moose breaks $G_L \times G_R$ into $G_{V'}$ which consists of $H_V$ and $(G/H)_A$. 

\section{Higgs Parity Operator for Symmetric Coset Spaces}
\label{App:Higgsparity}
In this Appendix we present the explicit form of the Higgs parity operator $V$ for various symmetric coset spaces. For $SU(M+N)/SU(M)\times SU(N) \times U(1)$ ($N \neq 1$ and $M \neq 1$) or  $SU(M+1)/SU(M)\times U(1)$  type of breaking, the fundamental representation can be decomposed as $(M+N) \to M_{1} + N_{-1}$, where the lower index $\pm 1$ is the $V$ parity. The adjoint of $SU(M+N)$ can be decomposed as the $(M_{1} +N_{-1}) \times (\bar{M}_{1} + \bar{N}_{-1}) = (M^2-1)_{1} + (N^2-1)_{1} + (\bar{M}\times N)_{-1} + (\bar{N}\times M)_{-1} +{\bf 1}_1$. Thus the broken generators have negative $V$ parity, while the unbroken ones positive, proving    that $V$ is the automorphism map of this symmetric space and is of the form diag$(1,1,...,-1)$. 
Similarly for $SO(M+1)/SO(M)$ or $SO(M+N)/SO(M) \times SO(N)$ spaces the automorphism map has the same form as above.

For $SU(2N)/Sp(2N)$, the VEV responsible for the breaking pattern and consequently also the Higgs parity operator $\Phi$ is an antisymmetric matrix belonging to the $SU(2N)$ group. The unbroken generators $T^{a}$ and  broken generators $T^{\hat{a}}$ satisfy~\cite{Galloway:2010bp}
\bea \label{eq:generator}
T^{a} \Phi + \Phi(T^{a})^T =0 \quad T^{\hat{a}} \Phi - \Phi(T^{ \hat{a} })^T =0 
\eea   
Thus the automorphism is given by 
\bea
T \to -\Phi T^T \Phi^\dagger  \Rightarrow U \to  \Phi U^\ast \Phi^\dagger
\eea
while the linearly realized sigma field, $\Sigma'$, and its transformation under the global $SU(2N)$ symmetry is
\bea \label{eq:sigma}
\Sigma &=& U \Phi U^T \Phi^\dagger = U^2 \quad  \Sigma \to L  \Sigma \Phi L^T \Phi^\dagger  \nonumber  \\
\Sigma' &=& \Sigma \Phi \Rightarrow \Sigma' \to L \Sigma' L^T \quad L \in  SU(2N)\ .
\eea
We can choose a basis where $\Phi$ is represented as 
\bea
\Phi ={\bf 1}_{N\times N} \times (i \sigma_2) 
\eea 
Similarly for  $SU(N)/SO(N)$, $\Phi$ is a symmetric matrix belonging to the $SU(N)$ group. The unbroken and broken generators satisfy the same relation as in Eq.~(\ref{eq:generator}). So the linear realized sigma field $\Sigma'$ is the same as the one in Eq.~(\ref{eq:sigma}). 
With an appropriate choice of basis  $\Phi$ can be written in the form of 
\bea
\Phi &=& \left(\begin{array}{cc}
0 & 1_{\frac{N}{2} \times \frac{N}{2} } \\  
1_{\frac{N}{2} \times \frac{N}{2} } & 0
\end{array} \right)   \quad  N = 2l \nonumber \\    
\Phi &=& \left(\begin{array}{ccc}
0 & & 1_{\frac{N-1}{2} \times \frac{N-1}{2} } \\
0 & 1 & 0 \\  
1_{\frac{N-1}{2} \times \frac{N-1}{2} } & & 0
\end{array} \right)   \quad  N = 2l +1 \nonumber \\  
\eea

\section{Symmetry Breaking Patterns in The Gauge Boson Sector for $SO(5)/SO(4)$}
\label{App:gauge}
 
For the $SO(5)/SO(4)$ MCH model the quantum numbers the of vector and axial-vector resonances $\rho_\mu$ and $a_\mu$ under $H =SO(4)$ are $\rho_\mu \equiv  {\bf 6}$ and $a_\mu \equiv  {\bf 4}$. In the hidden local symmetry approach~\cite{Bando:1987br}, under a global $SO(5)$ transformation $g$, these resonances transform as
\bea
\rho_\mu &\to & h \rho_\mu  h^\dagger + \frac{i}{g_\rho} h \partial_\mu h^\dagger \nonumber \\
a_\mu &\to &  h a_\mu h^\dagger 
\eea  
where ${ h }={ h}(g,h^{\hat{a}}) \in SO(4)$.
At leading order in derivatives, the most general Lagrangian can be written as (we assume for now only one copy of vector and axial resonances)~\cite{Marzocca:2012zn, Bian:2015ota}  
\bea
\mathcal{L}^{v} &=& -\frac{1}{4}\mbox{ Tr}[\rho_{\mu \nu} \rho^{\mu \nu}] +\frac{f_\rho ^2}{2} \mbox{Tr}[(g_{\rho} \rho_\mu -E_\mu )^2] \nonumber \\
\mathcal{L}^a &=& -\frac{1}{4}\mbox{ Tr}[a_{\mu \nu} a^{\mu \nu}] +\frac{f_a ^2}{2\Delta^2} \mbox{Tr}[(g_{a} a_\mu -\Delta d_\mu )^2] \nonumber \\
\mathcal{L}_{\mbox{kin}} &=& \frac{f^2}{4} \mbox{Tr}[d_\mu d^\mu],
\eea
where $U^\dagger D_\mu U =E_\mu ^a T^a +d_\mu ^{\hat{a}} T^{\hat{a}}$ and $T^{\hat{a}}$($T^{a}$) are (un)-broken generators and $D_\mu = \partial_\mu -ig_0 A_\mu ^a T^a$ is the gauge covariant derivative.
The field strengths and covariant derivatives are defined as 
\bea
 \rho_{\mu \nu} &=& \partial_\mu \rho_\nu -\partial_\nu \rho_\mu -i g_\rho [\rho_\mu, \rho_\nu], \nonumber \\
  a_{\mu \nu} &=& \nabla_\mu a_\nu -\nabla_\nu a_\mu, \quad \nabla =\partial -i E. 
 \eea 
Since the pNGB potential is generated from the mixing and the kinetic terms, we suppress the field strengths in following discusion. The total Lagrangian can be rewritten as 
\bea \label{eq:gauge}
 \mathcal{L} &=&  f_+ ^2 \mbox{Tr}[(d_\mu  + E_\mu )^2] 
 + f_- ^2 \mbox{Tr}[V( E_\mu +d_\mu)V (E^\mu+d_\mu)] \nonumber \\
 &-& m_+ ^2 \mbox{Tr}[(\rho_\mu +a_\mu )(d_\mu  + E_\mu )] \nonumber \\
 &-& m_- ^2 \mbox{Tr}[V(\rho_\mu +a_\mu )V(d_\mu  + E_\mu )] \nonumber \\
 &+& \frac{m^2_\rho +m_a^2 }{4}  \mbox{Tr}[ (\rho_\mu +a_\mu) (\rho_\mu+ a_\mu) ] \nonumber\\
& +&\frac{m^2_\rho -m_a^2 }{4}  \mbox{Tr}[ V(\rho_\mu +a_\mu) V(\rho_\mu+ a_\mu) ]
\eea 
 where $f_+ ^2 =\frac{f^2 +2f_a ^2 +2 f_\rho ^2}{8}$, $f_- ^2 =\frac{2 f_\rho ^2 -f^2 -2f_a ^2}{8}$, $m_+ ^2 =\frac{m_\rho f_\rho +m_a f_a}{2} $, $m_- ^2 =\frac{m_\rho f_\rho -m_a f_a}{2}$,  $m^2 _\rho =g_\rho ^2 f_\rho ^2$, $m_a ^2 = \frac{g_a ^2 f_a ^2}{\Delta^2}$.

We can see that the symmetry structure is very similar to the case of the fermion Lagrangian in~(\ref{eq:fulleq}).  We have the original shift symmetry on the pNGB's in $E_\mu +d_\mu =U^\dagger D_\mu U$ contained in an $SO(5)_1$ group: 
\bea
 U^\dagger D_\mu U \to \Omega_1  U^\dagger D_\mu U \Omega_1^\dagger 
\eea 
In addition, since $\rho_\mu +a_\mu $ can form a full adjoint representation of $SO(5)$, we can 
combine $\rho_\mu$ and $a_\mu$ to tranform under an additional $SO(5)_2$ as 
\bea
\rho_\mu +a_\mu \to \Omega_2 (\rho_\mu +a_\mu) \Omega_2^\dagger
\eea 
where $\Omega_{1,2}$ are $SO(5)_{1,2}$ transformations. Hence we have an enhanced $SO(5)_1\times SO(5)_2$ symmetry, which has various symmetry breaking patterns depending on the structure of the terms that are turned on in (\ref{eq:gauge}).   Just like for the fermion sector, the symmetry breaking patterns will determine the properties of the resulting induced pNGB potential. The main difference is that the analog of the maximal $SO(5)_{V'}$ symmetry in the gauge sector can not be achieved for physical parameters: it would correspond to $m_\rho^2+m_a^2=0$. Nevertheless the potential can be finite, and we will find the condition for a finite gauge contribution. First we summarize the main possibilities for the symmetry breaking patterns. 
 \begin{itemize}
\item  Consider first turning on only the $f_+ ^2$ and $f_- ^2$ terms. The $f_+$ term   is $SO(5)_1$ invariant so it does not contribute to the pNGB potential, while the  $f_-$ term necessarily breaks the  global symmetry to the $SO(4)_1$ subgroup.  Therefore the leading gauge contribution to the pNGB potential will be quadratically divergent and given by 
\bea
V_g \sim g_0^2 f_- ^2 \Lambda^2
\eea 
\item Next consider turning on only the $m_+ ^2$ and $m_- ^2$ terms. If $m_+ ^2 =0$, the $m_- ^2$ term  breaks the global $SO(5)_1 \times SO(5)_2$ into the maximal symmetry of the gauge sector $SO(5)_{D'}$ whose transformation is $\Omega_1 V \Omega_2 ^\dagger =V$. Since this unbroken group contains the pNGB shift symmetry the pNGB potential vanishes. If $m_- ^2=0$, for the same reason, the $m_+ ^2$ term  breaks the global $SO(5)_1 \times SO(5)_2$ into the the diagonal subgroup $SO(5)_{D}$ whose transformation is $\Omega_1 \Omega_2 ^\dagger =1$. This subgroup also contains the pNGB shift symmetry. But if $m_+ ^2 m_- ^2 \neq 0$, only the $SO(4)_D \in SO(5)_D$ subgroup is unbroken which does not contain the Higgs Goldstone symmetry. So the leading order of pNGB potential from these terms is proportional to 
\bea
V_g \sim g_0 ^2 m_+ ^2 m_- ^2 \mbox{log} \Lambda^2  
\eea
\item Now consider turning on only $m_+ ^2$ as well as the vector meson mass terms, $m_\rho^2 \neq 0$ and $m_a^2 \neq 0$. The mass term proportional to $m_\rho^2 +m_a ^2$ is $SO(5)_2$ invariant so this term does not contribute to the pNGB potential. However the term proportional to $m_\rho^2 -m_a ^2$ is only $SO(4)_2 \in SO(5)_2$ invariant. So if $m_\rho^2 -m_a ^2 =0$, $SO(5)_D$ is unbroken, the pNGB potential vanishes and thus the leading contribution to the potential is 
\bea
V_g \sim g_0 ^2  m_+ ^4(m_\rho^2 -m_a ^2)/ \Lambda^2  
\eea 
\item Similarly, if we only turn on $m_- ^2$ and the vector meson mass terms, the pNGB potential will be
\bea
V_g \sim g_0 ^2  m_- ^4(m_\rho^2 -m_a ^2)/ \Lambda^2  
\eea 
\end{itemize} 
So the pNGB potential from vector boson loops vanishes if and only if the global symmetry in the gauge sector is maximized, corresponding to the $SO(5)_D$ or $SO(5)_{D'}$ global symmetry, the conditions for which are given by 
\bea
 f^2 +2f_a ^2=2f_\rho ^2 \quad m_a^2 f_a^2 -m_\rho^2 f_\rho ^2 =0 \quad m_\rho ^2 -m_a ^2=0 \nonumber \\
\eea 
However these equations only have one solution $f^2=0, f_a^2=f_\rho^2, m_a^2=m_\rho^2$, and the limit of maximal symmetry is never realized. In any other case the gauge sector will contribute to the pNGB potential if there is only one copy of vector resonances.  If $m_\rho ^2 -m_a ^2 \neq 0$ (but the first two sum rules are still satisfied) the full global symmetry is collectively broken to $SO(4)$ and the pNGB potential is finite at one loop.  
If there are $N$ copies of vector mesons it is possible for the unbroken global symmetry to remain the maximal $SO(5)_V$ or $SO(5)_{V'}$ in which case the  gauge sector does not contribute to pNGB potential and the $S$-parameter also vanishes. The sum rule corresponding to a finite pNGB potential can also be easily generalized to the case with $N$ copies of the massive gauge boson sector:
\bea
 f^2 +\sum_i ^N 2f_a ^{i2}=\sum_i ^N 2f_\rho ^{i2} \quad  \sum_i m_a ^{i2} f_a ^{i2} =\sum_i m_\rho ^{i2} f_\rho ^{i2} \ . \nonumber \\
\eea

\section{Comparison to deconstructed models}
\label{App:deconstruction}
 
For models based on extra dimensions or their deconstructed version with a finite or log divergent potential, the higgs matrix $U$ is either given by the Wilson line or is the product of the several link fields. In this case, the massless (elementary) SM fermions are localized at the first site while all the composite fermions are localized at other sites with their own vectorlike masses. The global symmetry at every site is the full $G$,  except for the last site where this symmetry is spontaneously broken to $H$. For the case with maximal symmetry the vector fermion has a twisted mass with the Higgs parity operator $V$ inserted at the last site. By integrating out the heavy fields at the intermediate sites (or in the bulk), one can obtain the effective Lagrangian, the analog of Eq.~(\ref{eq:Leff}) for more sites. The leading divergent term in the Coleman-Weinberg potential is $\sim \int d^4p {\Pi_1^{q,t}}$, therefore the divergence of the Higgs potential is 4+div($\Pi_1^{q,t}$). For an N-site moose model the form factor of SM top kinetic terms $\slashed p \Pi^{q,t} _1$  contains $2N -3$ fermion propagators bvz, integrating out the composite fermions. If the Higgs parity operator $V$ twists the vectorlike fermion masses at the last site, then the form factor will be proportional to $1/(p^2-m^2_S) - 1/(p^2-m^2_Q)$ after integrating out the composite fermion at the last site, which implies an additional $p^{-2}$ suppression of the form factor. Therefore at large momenta $\slashed p \Pi^{q,t} _1$ are at least suppressed as $\mathcal{O}(p^{-2N+1})$ i.e. $\Pi_1 ^{q,t} \varpropto p^{-2N}$, implying that the  Higgs potential is finite for more than three sites. 

\section{Form factors}
\label{App:formfactors}
Here we present the explicit expressions of the form factor (\ref{eq:Leff}) obtained from integrating out the heavy fermions from the Lagrangian (\ref{eq:fulleq}). 
\bea
 \frac{\Pi_{0} ^{q,t}}{\lambda_{L,R} ^2 f^2} &=& 1+ \frac{(c_{-L,R}^2 + c_{+L,R}^2 )(M_Q ^2 +M_S ^2 -2 p^2)}{2(p^2 -M_S ^2)(M_Q ^2 -p^2)}  \nonumber \\
 +&& \frac{c_{-L,R} c_{+L,R}(M_S +M_Q)(M_S-M_Q)}{(p^2 -M_S ^2)(M_Q ^2-p^2) }  \nonumber \\     
 \frac{\Pi_{1} ^{q,t}}{\lambda_{L,R} ^2 f^2} &=& \frac{ c_{+L,R} c_{-L,R}(M_Q ^2 +M_S^2 -2p^2)}{(p^2 -M_S ^2)(M_Q ^2-p^2)}  \nonumber \\ 
 &+&\frac{ ( c_{+L,R}^2 +c_{-L,R}^2) (M_S-M_Q)(M_S +M_Q)}{2(p^2 -M_S ^2)(M_Q ^2-p^2) } \nonumber \\
\frac{ M_1 ^t }{\lambda_L \lambda_R  f^2} & = & \frac{M_Q ^2 M_S (c_{-L} - c_{+L})(c_{-R}-c_{+R})}{2(p^2 -M_Q^2)(p^2 -M_S ^2)}  \nonumber \\ 
 &-& \frac{ M_S ^2 M_Q (c_{-L} + c_{+L})(c_{-R} + c_{+R})}{2(p^2 -M_Q^2)(p^2 -M_S ^2)} \nonumber \\
&+& \frac{ M_Q (c_{-L} + c_{+L})(c_{-R} + c_{+R}) p^2}{2(p^2 -M_Q^2)(p^2 -M_S ^2)} \nonumber \\
&-&  \frac{ M_S (c_{-L} - c_{+L})(c_{-R} - c_{+R}) p^2}{2(p^2 -M_Q^2)(p^2 -M_S ^2)}  ,
\eea
 
\section{Numerical Scan}
\label{app:san}
\begin{figure}
\begin{center}
\includegraphics[width=8.7cm]
{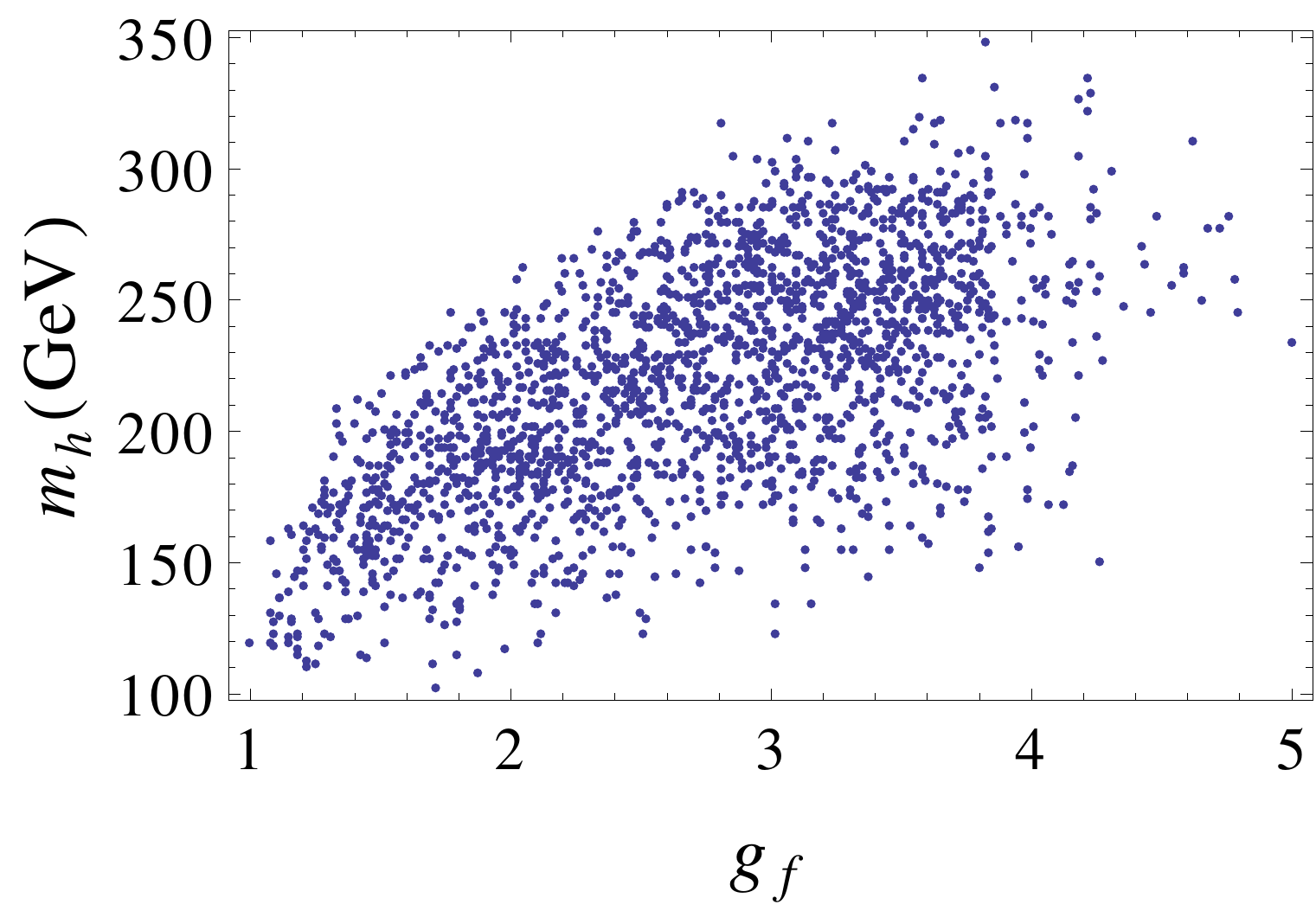}

\includegraphics[width=8.7cm]
{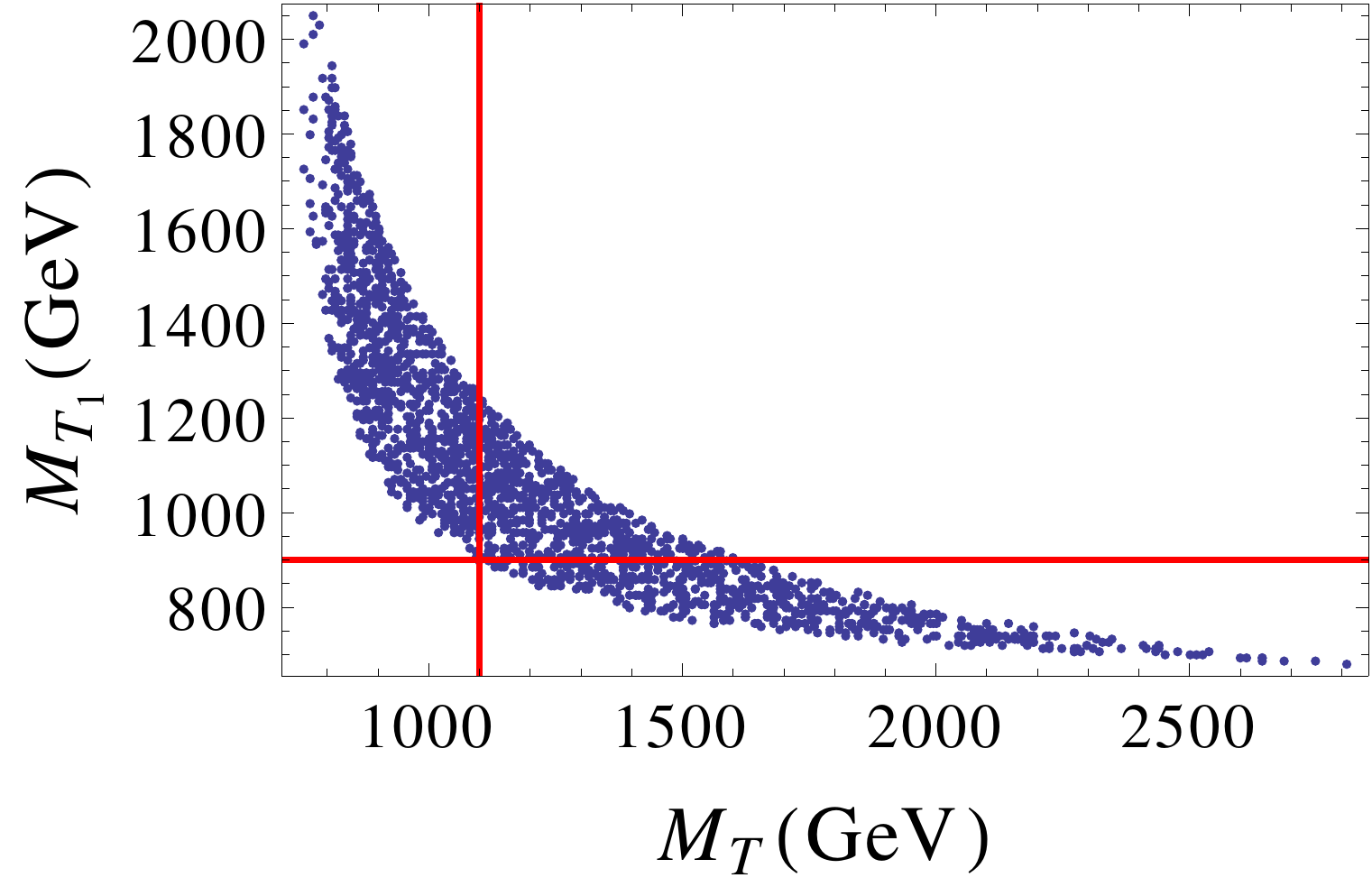}
\end{center}
\caption{Scatter plot in the MCHM$_5$ set-up for $\xi=0.1$. The range of the parameters is taken as follows: $m_t \in [150, 170]$ GeV, $m_\rho \geq 2$ TeV. In the top panel we show the Higgs mass as a function of $g_f$ and in the bottom panel the correlation of the mass of the doublet and singlet top partners for $m_h =125$ GeV. The horizontal and vertical red lines, corresponding to $900$ and $1100$~GeV respectively, are the lower bounds of the doublet and singlet top partners from the most recent $13$ TeV LHC data~\cite{ATLAS:2016btu, ATLAS:2016qlg,ATLAS:2016cuv}.} 
\label{fig:CHM}  
\end{figure}

We show scatter plots for the maximally symmetric MCHM$_5$ set-up for $\xi=0.1$ in Fig.~\ref{fig:CHM}. The range of the parameters is taken as follows: $m_t \in [150, 170]$ GeV and $m_\rho \geq 2$ TeV. In the top panel the Higgs mass as a function of $g_f$ and in the bottom panel the correlation of the doublet and singlet top partner mass, $M_T$ and $M_{T_1}$  for $m_h=125$ GeV. The horizontal and vertical red lines, corresponding to $900$ and $1100$~GeV respectively, are the lower limits of the doublet and singlet top partners from $13$ TeV LHC (from 13.2 to 14.7 fb$^{-1}$ data)~\cite{ATLAS:2016btu, ATLAS:2016qlg,ATLAS:2016cuv}\footnote{The LHC Run 2 bounds on an exotic charge 5/3 fermion has not been published yet,  we expect that the updated result will give us a much stronger bound.}. 
 
\end{document}